\documentclass[11pt,a4paper]{article}

\usepackage{fullpage}
\usepackage{amsmath}
\usepackage{amssymb}
\usepackage{mathrsfs}
\usepackage{graphicx}
\usepackage{psfrag}
\usepackage[all,poly,curve,knot,cmtip]{xy}
\usepackage{citesort}

\newcommand {\be} {\begin{eqnarray*}}
\newcommand {\ee} {\end{eqnarray*}}
\newcommand {\bea} {\begin{eqnarray}}
\newcommand {\eea} {\end{eqnarray}}

\newcommand{\bm}[1]{\boldsymbol{#1}}

\newcommand{\pdiff}[2]{\frac{\partial{#1}}{\partial{#2}}}

\begin{document}

\title{Isolated horizons, $p$-form matter fields, topology and the black-hole/string
correspondence principle}
\author{\textbf{Tom$\acute{\mbox{a}}\check{\mbox{s}}$
Liko}\footnote{Electronic mail: liko@gravity.psu.edu}\\
\\{\small \it Institute for Gravitation and the Cosmos}\\
{\small \it Pennsylvania State University}\\
{\small \it University Park, Pennsylvania 16802, USA}}

\maketitle

\begin{abstract}

We study the mechanics of $D$-dimensional isolated horizons (IHs) for Einstein
gravity in the presence of arbitrary $p$-form matter fields.  This generalizes
the analysis of Copsey and Horowitz to non-stationary spacetimes and therefore
the local first law in $D>4$ dimensions to include non-monopolar (dipole)
charges.  The only requirement for the local first law to hold is that the
action has to be differentiable.  The resulting conserved charges are all
intrinsic to the horizon and are independent of the topology of the horizon
cross sections.  We explicitly calculate the local charges for
five-dimensional black holes and black rings that are relevant within the
context of superstring theory.  We conclude with some comments on the
black-hole/string correspondence principle and argue that IHs (or some other
quasilocal variant) should play a fundamental role in superstring theory.

\end{abstract}

\hspace{0.35cm}{\small \textbf{PACS}: 04.50.Gh; 04.70.Bw}



\section{Introduction}

The advent of superstring theory revolutionized our view of the universe, for example, with the
requirement of extra spatial dimensions.
%
%
The natural question that should be investigated is the following: \emph{What properties
of black holes in four dimensions carry over to higher-dimensional spacetimes?}  More specifically,
we should ask the following question: \emph{What are the generic features of black holes in
higher-dimensional spacetimes in general, and within the superstring theory context in particular?}
An ideal method for investigating such questions is to employ a covariant phase space framework that
includes all black-hole solutions to the equations of motion for a given action principle.

Such a framework does exist and is known as the isolated horizon (IH) framework \cite{abdfklw}.  The
classical theory of IHs was motivated by earlier considerations of trapping horizons \cite{hayward},
but the framework is considerably different as covariant phase space methods
\cite{leewal,abr,walzou,ashstr} are employed in the case of IHs.  All the quantities that appear in
the first law of IH mechanics are defined intrinsically at the horizon.  The concept of such a
surface generalizes the notion of a Killing horizon to much more general and therefore physical
spacetimes that may include external radiation fields that are dynamical.  Examples of such systems
in general relativity are given by the so-called Robinson-Trautman spacetimes \cite{abf1,lewandowski}.

The focus of this paper is to examine the consequences of the IH boundary conditions on the covariant
phase space of solutions to the equations of motion in the presence of generic $p$-form matter fields
and to determine the conserved charges from the symplectic structure.  Among other results, we find
that the natural conserved charge associated with the matter term for the electric dual of
Einstein-Maxwell theory with dilaton that arises from the symplectic structure is the electric dipole
charge, not the magnetic monopolar charge that one would expect.  This work generalizes two sets of
constructions: the first law of Copsey and Horowitz \cite{cophor} is generalized to non-stationary
spacetimes and the IH framework in $D>4$ dimensions \cite{klp,apv,likboo} is extended to include
non-monopolar charges.

We consider a $D$-dimensional manifold $\mathcal{M}$ bounded by two spacelike partial Cauchy surfaces,
$M_1$ and $M_2$, which are asymptotically related by a time translation and extend from the internal
boundary $\Delta$ [with $\Delta\cap M\cong\mathbb{S}^{D-2}$ for some compact $(D-2)$-space
$\mathbb{S}^{D-2}$ with positive constant curvature] to the boundary at infinity $\tau_{\infty}$.
See Figure 1.

In the first-order formulation of general relativity the action for the theory that we consider is given
by
\bea
S = \frac{1}{2\kappa_{D}}\int_{\mathcal{M}}\Sigma_{IJ} \wedge \Omega^{IJ}
    + \mathcal{L}_{\rm M}[\Phi,\bm{\mathcal{F}};\bm{\mathcal{A}}]
    - \frac{1}{2\kappa_{D}}\int_{\tau_{\infty}}\Sigma_{IJ} \wedge A^{IJ} \; .
\label{action}
\eea
Here, $\kappa_{D}=8\pi G_{D}$ with $G_{D}$ the $D$-dimensional gravitational constant.  This action depends
on the co-frame $e^{I}$, the gravitational $SO(D-1,1)$ connection $A_{\phantom{a}J}^{I}$, the scalar field
$\Phi$ and the generic $p$-form field $\bm{\mathcal{F}}=d\bm{\mathcal{A}}$ (with $p$ an integer such that
$2\leq p \leq D-2$).
The co-frame determines the metric $g_{ab}=\eta_{IJ}e_{a}^{\phantom{a}I} \otimes e_{b}^{\phantom{a}J}$,
$(D-2)$-form $\Sigma_{IJ}=[1/(D-2)!]\epsilon_{IJK_{1}\ldots K_{D-2}}e^{K_{1}} \wedge \cdots \wedge e^{K_{D-2}}$
and spacetime volume form $\bm{\epsilon}=e^{0} \wedge \cdots \wedge e^{D-1}$, where $\epsilon_{I_{1}\ldots I_{D}}$
is the totally antisymmetric Levi-Civita tensor.  The connection determines the curvature two-form
\bea
\Omega_{\phantom{a}J}^{I} = dA_{\phantom{a}J}^{I}
+A_{\phantom{a}K}^{I} \wedge A_{\phantom{a}J}^{K}
= \frac{1}{2}R_{\phantom{a}JKL}^{I}e^{K} \wedge e^{L} \, ,
\eea
with $R_{\phantom{a}JKL}^{I}$ as the Riemann tensor.  In this paper, spacetime indices
$a,b,\ldots\in\{0,\ldots,D-1\}$ are raised and lowered using the metric $g_{ab}$ and internal indices
$I,J,\ldots\in\{0,\ldots,D-1\}$ are raised and lowered using the Minkowski metric
$\eta_{IJ}=\mbox{diag}(-1,\ldots,1)$.  The boundary term at the timelike cylinder $\tau_{\infty}$ at
infinity is required in order that the action be differentiable.  It is the natural boundary term associated
with the first-order action principle.
Important properties of this boundary term are discussed in \cite{aes,ashslo,likslo}.
%
\begin{figure}[t]
\begin{center}
\psfrag{D}{$\Delta$}
\psfrag{Mp}{$M_{2}$}
\psfrag{Mm}{$M_{1}$}
\psfrag{Mi}{$M$}
\psfrag{Sp}{$\mathbb{S}_{2}$}
\psfrag{Sm}{$\mathbb{S}_{1}$}
\psfrag{S}{$\mathbb{S}$}
\psfrag{B}{$\tau_{\infty}$}
\psfrag{M}{$\mathcal{M}$}
\includegraphics[width=4.5in]{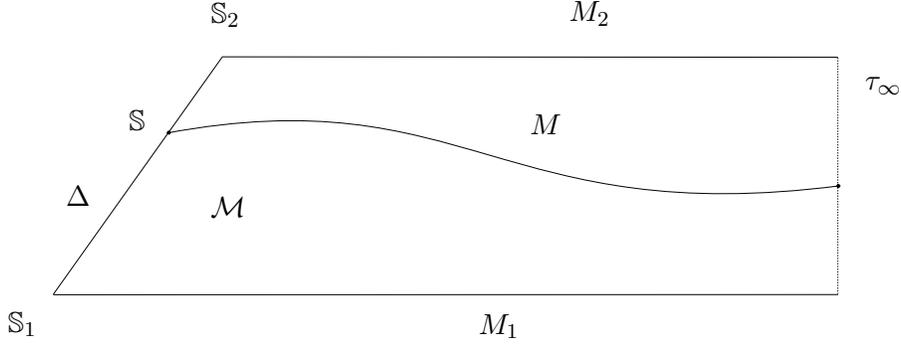}
\caption{The spacetime manifold $\mathcal{M}$ and its boundary
$\partial\mathcal{M}\cong M_{1}\cup M_{2}\cup\Delta\cup\tau_{\infty}$.  Here, $M_1$ and $M_2$ are partial
Cauchy surfaces that are asymptotically related by a time translation, and extend from the internal boundary
$\Delta$ to the boundary at infinity $\tau_{\infty}$.  The partial Cauchy surface $M$ intersects $\Delta$ in
a compact $(D-2)$-space $\mathbb{S}$ with positive constant curvature.}
\end{center}
\end{figure}

\section{Horizon structures and differentiability of the action}

Let us remind the reader of the basic definition of a rotating weakly isolated horizon \cite{afk,abl,apv},
with a suitable generalization of the boundary conditions tailored to include the presence of $p$-form matter
fields.

\noindent{\bf Definition I.}
\emph{A rotating weakly isolated horizon (WIH) $\Delta$ is a null surface and has a degenerate metric $q_{ab}$
with signature $0+\ldots+$ (with $D-2$ non-degenerate spatial directions) along with an equivalence class of
null normals $[\ell]$ (the equivalence relation being defined by $\ell^{\prime}=z\ell$ for some constant $z$)
and spacelike rotational vector fields $\phi_{\iota}^{a}$ ($\iota\in\{1,\ldots,\lfloor (D-1)/2 \rfloor |\lfloor
\cdot \rfloor~\mbox{denotes ``integer value of''}\}$) such that the following conditions hold: (1) the expansion
$\theta_{(\ell)}$ of $\ell_{a}$ vanishes on $\Delta$; (2) the field equations hold on $\Delta$; (3) the
stress-energy tensor is such that the vector $-T_{\phantom{a}b}^{a}\ell^{b}$ is a future-directed and causal
vector; (4) $\pounds_{\ell}\omega_{a}=0$ and $\pounds_{\ell}\underleftarrow{\bm{\mathcal{A}}}=0$ for all
$\ell\in[\ell]$ (see below); (5) $\phi_{\iota}^{a}$ satisfy $\pounds_{\phi}q_{ab}=\pounds_{\phi}\ell_{a}
=\pounds_{\phi}\omega_{a}=\pounds_{\phi}\underleftarrow{\bm{\mathcal{A}}}
=\pounds_{\phi}\underleftarrow{\bm{\mathcal{F}}}=0$.}

The first three conditions determine the intrinsic geometry of $\Delta$. Since $\ell$ is normal to 
$\Delta$ the associated null congruence is necessarily twist-free and geodesic. By condition (1) that congruence is
non-expanding. Then the Raychaudhuri equation implies that $T_{ab}\ell^{a}\ell^{b}=-\sigma_{ab}\sigma^{ab}$, with
$\sigma_{ab}$ the shear tensor, and applying the energy condition (3) we find that $\sigma_{ab} = 0$.

In addition, the vanishing of the expansion, twist and shear imply that \cite{afk}
\bea
\nabla_{\!\underleftarrow{a}}\ell_{b}\approx\omega_{a}\ell_{b} \, ,
\label{connectionondelta}
\eea
with ``$\approx$'' denoting equality restricted to $\Delta$ and the underarrow indicating pull-back to $\Delta$.
Thus the one-form $\omega$ is the natural connection induced on the horizon.  The conditions also imply that
\bea
\underleftarrow{\ell \lrcorner \bm{\mathcal{F}}} = 0 \; .
\label{pullback1}
\eea
This property will play an important role in the derivation of the first law with non-monopolar charges for black
rings.  We emphasize that this condition is a consequence of the boundary conditions and not an assumption.

Condition (5) captures the notion of a WIH rotating with angular velocities $\Omega_{\iota}$ whereby the
rotational vector fields $\phi_{\iota}^{a}$ are symmetries of $\Delta$.  For a multidimensional rotating
WIH, a suitable evolution vector field on the covariant phase space is given by \cite{abl,apv}
\bea
\xi^{a} = z\ell^{a} + \sum_{\iota=1}^{\lfloor (D-1)/2 \rfloor}\Omega_{\iota}\phi_{\iota}^{a} \; .
\eea
This vector field is spacelike in general and becomes null when all angular momenta are zero.

We do not fix the fields at the inner boundary $\Delta$, so we need to determine explicitly the surface
terms for which the action (\ref{action}) will be differentiable.  To this end, let
$\Psi\in\{e,A,\Phi,\bm{\mathcal{F}}$\} denote the set of field variables.  Then, taking the first variation
of (\ref{action}) gives
\bea
\delta S = \frac{1}{2\kappa_{D}}\int_{\mathcal{M}}E[\Psi]\delta\Psi - \frac{1}{2\kappa_{D}}\int_{\Delta}J[\Psi,\delta\Psi] \, ,
\eea
with $E[\Psi]=0$ representing the equations of motion and $J[\Psi,\delta\Psi]$ representing a linear combination
of gravitational and matter-field surface terms.  In the present case, we have that
\bea
J[\Psi,\delta\Psi] = \Sigma_{IJ} \wedge \delta A^{IJ}
                     + \bm{\Upsilon} \wedge \delta\bm{\mathcal{A}} \, ;
\label{surface}
\eea
here we defined $\bm{\Upsilon}=\mathbb{D}\mathcal{L}_{\rm M}/\mathbb{D}\bm{\mathcal{F}}$ as the functional
derivative of the Lagrangian density $\mathcal{L}_{\rm M}$ with respect to $\bm{\mathcal{F}}$.

It turns out that the pull-back of $J$ to $\Delta$ vanishes, and therefore the action (\ref{action}) is indeed
differentiable and the equations of motion $E[\Psi]=0$ follow from the variational principle $\delta S=0$.  In
particular, the pull-back of the gravitational surface term is given by \cite{likboo}
\bea
\underleftarrow{\Sigma \wedge \delta A} \approx \bm{\tilde{\epsilon}} \wedge \delta \omega \, ,
\eea
with $\bm{\tilde{\epsilon}}=\vartheta^{(1)} \wedge \dots \wedge \vartheta^{(D-2)}$ the area element of the
cross section $\mathbb{S}^{D-2}$ of the horizon, and $\vartheta_{(i)}$ ($i\in\{2,\ldots,D-1\}$) are $D-2$
spacelike vectors adapted to $\mathbb{S}^{D-2}$ that satisfy the orthogonality condition
$\vartheta_{(i)} \cdot \vartheta_{(j)}=\delta_{ij}$.  The key property of $\Delta$ is that the variation
of $\ell$ is proprtional to $\ell$ itself.  Then from the WIH condition (4) it follows that
$\pounds_{\ell}\delta\omega=0$.  However, $\omega$ is held fixed on $M_{\{1,2\}}$ which means that
$\delta\omega=0$ on the initial and final cross-sections of $\Delta$ (i.e. on $M_{1}\cap\Delta$ and on
$M_{2}\cap\Delta$), and because $\delta\omega$ is Lie dragged on $\Delta$ it follows that
$\bm{\tilde{\epsilon}} \wedge \delta \omega\approx0$.  The same argument also holds for the matter-field
part of the surface term: from condition (4) $\pounds_{\ell}\underleftarrow{\bm{\mathcal{A}}}=0$, and with
$\delta\ell\propto\ell$ on $\Delta$ it follows that
$\underleftarrow{\bm{\Upsilon} \wedge \delta\bm{\mathcal{A}}}\approx0$, whence
\bea
J[\Psi,\delta\Psi]|_{\Delta} \approx 0 \; .
\eea
Therefore in the presence of an internal null boundary $\Delta$ satisfying the conditions of Definition I,
the action (\ref{action}) is differentiable.

\section{Covariant phase space and conserved charges}

As in the previous papers on IHs (e.g. \cite{afk,abl,apv,likboo}), the first law follows directly from
applying standard covariant phase space methods \cite{leewal,abr,walzou,ashstr}.  The symplectic current
is obtained from antisymmetrizing the second variation of the surface term; integrating over the boundary
$M_{1}\cup M_{2}\cup \Delta$ (because the asymptotic conditions ensure that the integral over
$\tau_{\infty}$ vanishes) gives the symplectic structure
$\bm{\Omega}\equiv\bm{\Omega}(\delta_{1},\delta_{2})$.  The first law then follows directly from
evaluating the symplectic structure at $(\delta,\delta_{\xi})$.

In the present case the closed and conserved symplectic structure is given by
%
%
%
%
%
%
%
%
%
%
%
%
%
%
%
%
\bea
\bm{\Omega}(\delta_{1},\delta_{2})
&=& \frac{1}{2\kappa_{D}}\int_{M}
    \left[\delta_{[1}\Sigma_{IJ} \wedge \delta_{2]}A^{IJ}
    - \delta_{[1}\bm{\Upsilon} \wedge \delta_{2]}\bm{\mathcal{A}}\right]\nonumber\\
& & + \frac{1}{\kappa_{D}}\oint_{\mathbb{S}^{D-2}}
    \left[\delta_{[1}\bm{\tilde{\epsilon}} \wedge \delta_{2]}\psi
    + \delta_{[1}\bm{\Upsilon} \wedge \delta_{2]}\bm{\chi}\right] \; .
\label{fullsymplectic}
\eea
Here we defined the potential $\psi$ for the surface gravity $\kappa_{(\ell)}$ and $(p-2)$-form $\bm{\chi}$
for the $(p-2)$-form $\bm{\Phi}_{(\ell)}$ such that
\bea
\pounds_{\ell}\psi \approx \ell \lrcorner \omega = \kappa_{(\ell)}
\quad
\mbox{and}
\quad
\pounds_{\ell}\bm{\chi} \approx \ell \lrcorner \bm{\mathcal{A}} = -\bm{\Phi}_{(\ell)} \, ,
\label{potentials}
\eea
We find that evaluating the horizon integral at $(\delta,\delta_{\xi})$ is given by
\bea
\bm{\Omega}|_{\Delta}
&=& \frac{1}{\kappa_{D}}\oint_{\mathbb{S}^{D-2}}\kappa_{(z\ell)}\delta\bm{\tilde{\epsilon}}
    + \frac{1}{\kappa_{D}}\oint_{\mathbb{S}^{D-2}}\bm{\Phi}_{(z\ell)} \wedge \delta\bm{\Upsilon}\nonumber\\
& & + \sum_{\iota=1}^{\lfloor (D-1)/2 \rfloor}\frac{\Omega_{\iota}}{\kappa_{D}}
    \delta\oint_{\mathbb{S}^{D-2}}\left[(\phi_{\iota} \lrcorner \omega)\bm{\tilde{\epsilon}}
    + (\phi_{\iota} \lrcorner \bm{\mathcal{A}}) \wedge \bm{\Upsilon}\right] \, ,
\label{evaluation}
\eea
where we used $\kappa_{(z\ell)}=\pounds_{z\ell}\psi=z\ell\lrcorner\omega$ and
$\bm{\Phi}_{(z\ell)}=\pounds_{z\ell}\bm{\chi}=z\ell\lrcorner\bm{\mathcal{A}}$.  These quantities are
constant for any given horizon, but in general vary across the phase space from one point to
another.  This implies that (\ref{evaluation}) is in general \emph{not} a total variation.
However, if $\kappa_{(z\ell)}$, $\bm{\Phi}_{(z\ell)}$ and $\Omega_{\iota}$ can be expressed as
functions of the entropy $\mathcal{S}$, charge $\mathcal{Q}$ and angular momenta
$\mathcal{J}_{\iota}$ defined by
\bea
\mathcal{S}     &=& \frac{1}{4G_{D}}\oint_{\mathbb{S}^{D-2}}\bm{\tilde{\epsilon}} \, ,
\label{entropy}\\
\mathcal{Q}     &=& \frac{1}{8\pi G_{D}}\oint_{\mathbb{S}^{D-2}}\bm{\Upsilon} \, ,
\label{charge}\\
\mathcal{J}_{\iota} &=& \frac{1}{8\pi G_{D}}\oint_{\mathbb{S}^{D-2}}
                    \left[(\phi_{\iota} \lrcorner \omega)\bm{\tilde{\epsilon}}
                    + (\phi_{\iota} \lrcorner \bm{\mathcal{A}}) \wedge \bm{\Upsilon}\right] \, ,
\label{angularmomentum}
\eea
and satisfy the integrability conditions
\bea
\pdiff{\kappa}{\mathcal{J}} = \pdiff{\Omega}{\mathcal{S}} \, ,
\quad
\pdiff{\kappa}{\mathcal{Q}} = \pdiff{\bm{\Phi}}{\mathcal{S}} \, ,
\quad
\pdiff{\Omega}{\mathcal{Q}} = \pdiff{\bm{\Phi}}{\mathcal{J}} \, ,
\eea
then there exists a function $\mathcal{E}$ such that \cite{abl,apv}
\bea
\bm{\Omega}|_{\Delta}(\delta,\delta_{\xi})=\delta \mathcal{E} \; .
\eea
In this case (\ref{evaluation}) becomes
\bea
\delta \mathcal{E} = \frac{\kappa_{(z\ell)}}{2\pi}\delta \mathcal{S}
                     + \frac{1}{\kappa_{D}}\oint_{\mathbb{S}^{D-2}}\bm{\Phi}_{(z\ell)} \wedge \delta\bm{\Upsilon}
                     + \sum_{\iota=1}^{\lfloor (D-1)/2 \rfloor}
                         \Omega_{\iota}\delta \mathcal{J}_{\iota} \, ,
\label{firstlaw1}
\eea
which is the first law (for a quasi-static process).  Therefore rotating WIHs in $D$-dimensional
asymptotically flat spacetimes with generic $p$-form matter fields satisfy the first law.

The first law (\ref{firstlaw1}) holds for any rotating WIH in the presence of $p$-form matter fields,
regardless of the topology of the horizon cross section.  For WIHs in asymptotically flat spacetimes,
there is a very strong constraint on the possible topologies.  As was shown in \cite{likboo}, the
integral of the scalar curvature of the horizon cross section is strictly positive.  This implies
that in four dimensions (together with the Gauss-Bonnet theorem) $\mathbb{S}^{2}\cong S^{2}$ and that
in five dimensions $\mathbb{S}^{3}$ can only be a finite connected sum of the three-sphere $S^{3}$ or
of the ring $S^{1}\times S^{2}$.  These results on topology are in agreement with the recent extension
of the Hawking topology theorem to higher dimensions \cite{caigal,galloway,galsch,hoy,racz}.

In addition, we note that the first law (\ref{firstlaw1}) is the \emph{equilibrium version} of the first
law of black-hole mechanics.  That is, (\ref{firstlaw1}) relates the infinitesimal changes in the conserved
charges of two nearby WIHs within the covariant phase space of solutions.  However, as was discussed in
\cite{abf2}, a local first law such as (\ref{firstlaw1}) also has a natural interpretation as the
\emph{physical process} version of black-hole mechanics \cite{gaowal,rogatko1,rogatko2}, whereby the
infinitesimal changes of the conserved charges of a single black hole are related when a small mass is
dropped into the horizon and the black hole is allowed to settle into a new equilibrium state.

An extension of the current framework to asymptotically anti-de Sitter (ADS) spacetimes, along the
lines of \cite{apv}, is straightforward.  In the presence of a negative cosmological constant
$\Lambda=-(D-1)(D-2)/(2L^{2})$ the covariant phase space of WIHs is modified to include a set of
conserved charges at the boundary at infinity $\mathscr{S}$ (with $\mathscr{S}\cap M\cong\mathbb{C}^{D-2}$
for some compact $(D-2)$-space $\mathbb{C}^{D-2}$) \cite{apv}.  These are the Ashtekar-Magnon-Das (AMD)
charges \cite{ashmag,ashdas}
\bea
\mathscr{Q}_{\xi}^{(\mathscr{I})} = \frac{L}{8\pi G_{D}}\oint_{\mathbb{C}^{D-2}}
                                    \widetilde{E}_{ab}k^{a}\tilde{u}^{b}\bm{\tilde{\varepsilon}} \, ,
\label{amdcharges}
\eea
with $k^{a}$ a Killing vector field that generates a symmetry (i.e. time translation etc),
$\tilde{u}^{a}$ the unit timelike normal to $\mathbb{C}^{D-2}$, $\bm{\tilde{\varepsilon}}$
the area form on $\mathbb{C}^{D-2}$ and $\widetilde{E}_{ab}$ the leading-order electric part
of the Weyl tensor $\widetilde{C}_{abcd}$.  Explicitly we have that
\bea
\widetilde{E}_{ab} = \frac{1}{D-3}\Omega^{3-D}\widetilde{C}_{abcd}\tilde{n}^{c}\tilde{n}^{d} \, ,
\eea
where $\tilde{n}^{a}=\tilde{\nabla}^{a}\Omega$ and $\Omega$ is a function on the conformally completed
manifold $\widehat{\mathcal{M}}\cong\mathcal{M}\cup\mathscr{S}$ that defines the unphysical metric $\tilde{g}_{ab}$
on $\mathcal{M}$ in terms of the physical spacetime metric $g_{ab}$ via $\tilde{g}_{ab}=\Omega^{2}g_{ab}$.
As was shown in Appendix B of \cite{him}, inclusion of antisymmetric tensor fields in the action does not
contribute anything to the charges at $\mathscr{I}$ because the fields fall off too quickly.  In particular
this implies that in the presence of generic $p$-form fields the charges at infinity are the AMD charges.
It is important to keep in mind that the charges at $\mathscr{I}$ are the charges of
the spacetime and are independent of the local charges at $\Delta$.

\section{Example theories}

The preceding analysis was rather abstract and technical.  In this section we will apply the framework to
two effective actions that arise within superstring theory.  This will serve to illustrate the generality
of the IH framework and will lead to some interesting surprises.

Let us consider first Einstein-Maxwell theory in five dimensions with electromagnetic Chern-Simons term.
The action for this theory in five dimensions is given by
\bea
S = \frac{1}{2\kappa_{5}}\int_{\mathcal{M}}\Sigma_{IJ} \wedge \Omega^{IJ}
    - \frac{1}{4}\bm{F} \wedge \star \bm{F}
    - \frac{2}{3\sqrt{3}}\bm{A} \wedge \bm{F} \wedge \bm{F}
    - \frac{1}{2\kappa_{5}}\int_{\tau_{\infty}}\Sigma_{IJ}\wedge A^{IJ} \; .\nonumber\\
\label{action2}
\eea
Here, $\bm{F}=d\bm{A}$ is the field strength of the connection one-form $\bm{A}$ and ``$\star$'' denotes
the Hodge dual.  The last term is a Chern-Simons (CS) term for the electromagnetic field.  For this theory
we take $\bm{\mathcal{F}}=\bm{F}$ and $\bm{\mathcal{A}}=\bm{A}$.  Then $\bm{\Phi}_{(z\ell)}=
-z\ell\lrcorner\bm{A}=\Phi_{(z\ell)}$ is just a scalar potential and
$\bm{\Upsilon}=\star\bm{F}+[4/(3\sqrt{3})]\bm{A} \wedge \bm{F}$.  The first law then takes the form
\bea
\delta \mathcal{E} = \frac{\kappa_{(z\ell)}}{2\pi}\delta \mathcal{S}
                     + \Phi_{(z\ell)}\delta\mathcal{Q}
                     + \sum_{\iota=1}^{\lfloor (D-1)/2 \rfloor}\Omega_{\iota}\delta \mathcal{J}_{\iota} \, ,
\label{firstlaw2}
\eea
with electric charge $\mathcal{Q}$ given by
\bea
\mathcal{Q} = \frac{1}{8\pi G_{5}}\oint_{\mathbb{S}^{3}}\star\bm{F} + \frac{4}{3\sqrt{3}}\bm{A} \wedge \bm{F} \; .
\label{charge1}
\eea
This is the natural conserved charge for both $\mathbb{S}^{3}\cong S^{3}$ and $\mathbb{S}^{3}\cong S^{1}\times S^{2}$
topologies; it is a monopolar electric charge.



Let us now consider the electric dual of Einstein-Maxwell theory with dilaton in five dimensions.  The action for this
theory is given by
\bea
S = \frac{1}{2\kappa_{5}}\int_{\mathcal{M}}\Sigma_{IJ} \wedge \Omega^{IJ}
    - \frac{1}{12}\mbox{e}^{-\alpha\varphi}\bm{H} \wedge \star \bm{H}
    - \frac{1}{2}d\varphi \wedge \star d\varphi
    - \frac{1}{2\kappa_{5}}\int_{\tau_{\infty}}\Sigma_{IJ}\wedge A^{IJ} \; .\nonumber\\
\label{action4}
\eea
Here, $\varphi$ is the dilaton field with coupling $\alpha$, and $\bm{H}=d\bm{B}$ is the field strength of the two-form
$\bm{B}$.  Because $\bm{H}$ is a three-form, one expects to \emph{define} a magnetic monopolar charge associated with
black holes within this theory.  However, this is not the case for IHs.  As we will now show, the IH boundary conditions
will give a dipolar electric charge that is conserved.  For this theory we take $\bm{\mathcal{F}}=\bm{H}$ and
$\bm{\mathcal{A}}=\bm{B}$.  Then $\bm{\Phi}_{(z\ell)}=-z\ell\lrcorner\bm{B}$ is a one-form potential, and
$\bm{\Upsilon}=\mbox{e}^{-\alpha\varphi}\star\bm{H}$.  The first law then takes the same form as (\ref{firstlaw2}), but
with a charge $\mathcal{Q}$ that is radically different from the electric charge (\ref{charge1}).  Here we have
\bea
\oint_{\mathbb{S}^{3}}\bm{\Phi}_{(z\ell)} \wedge \delta\bm{\Upsilon}
= \oint_{\mathbb{S}^{3}}(z\ell\lrcorner\bm{B}) \wedge \delta\left(\mbox{e}^{-\alpha\varphi}\star\bm{H}\right) \; .
\eea
The key observation is that $\bm{\Phi}_{(z\ell)}$ is a closed one-form \emph{at the horizon}.  This follows from the
Cartan identity $d(z\ell\lrcorner\bm{B})=\pounds_{z\ell}\bm{B}-z\ell\lrcorner d\bm{B}$; pulling this identity back to the
horizon gives
\bea
d(\underleftarrow{z\ell\lrcorner\bm{B}}) = \pounds_{z\ell}\underleftarrow{\bm{B}} - \underleftarrow{z\ell \lrcorner \bm{H}} \; .
\eea
Then from Condition (4) of Definition I and equation (\ref{pullback1}) it immediately follows that the right hand side is
zero.  Because $d(\underleftarrow{z\ell\lrcorner\bm{B}})\approx0$ we conclude that at the horizon $z\ell\lrcorner\bm{B}$ is
a closed one-form and must therefore be the sum of an exact one-form $df$ and harmonic one-form $dh$.  That is,
\bea
z\ell\lrcorner\bm{B} \approx df + cdh \, ,
\eea
with $c$ a constant.  The only non-zero contribution to the charge then comes from integrating $h$ over $S^{1}$, otherwise
the charge is zero \cite{cophor} (see also \cite{rogatko3,compere}).  Thus taking $2\pi$ to be the affine length of $S^{1}$,
we conclude that
\bea
\oint_{S^{1}\times S^{2}}cdh \wedge \delta\left(\mbox{e}^{-\alpha\varphi}\star\bm{H}\right)
= 2\pi c\delta\oint_{S^{2}}\mbox{e}^{-\alpha\varphi}\star\bm{H} \, ,
\eea
whence the charge
\bea
\mathcal{Q} = \frac{1}{8\pi G_{5}}\oint_{S^{2}}\mbox{e}^{-\alpha\varphi}\star\bm{H} \; .
\label{charge3}
\eea
This is the natural conserved charge for the $\mathbb{S}^{3}\cong S^{1}\times S^{2}$ topology.  By contrast to the previous
charge (\ref{charge1}), however, (\ref{charge3}) is a \emph{dipole} electric charge.

The first law (\ref{firstlaw1}) that we obtained for IHs is in agreement with that which was found for stationary spacetimes
\cite{cophor}.  However, we note that in the latter approach there also appeared a dipole charge in the first law for
Einstein-Maxwell theory with electromagnetic Chern-Simons term.  This charge does not appear in (\ref{firstlaw2}) which means
that the dipole charge, although possible to define, is not a \emph{conserved} charge for IHs.  This is in agreement with what
is known about the black ring solutions of Elvang \emph{et al} \cite{eemr1,eef,eemr2}.

The dipole charge (\ref{charge3}) that we obtained for the electric dual of Einstein-Maxwell theory with dilaton is in
agreement with that obtained for stationary spacetimes \cite{cophor} for the dipole ring solution \cite{emparan}.  There
the dipole charge is interpreted as an electric Kalb-Ramond charge localized on a fundamental string that winds around a
contractible circle \cite{astrad}.  However, the other conserved charges are still measured at infinity.  By contrast,
here we have found a first law whereby \emph{all} conserved charges, including the dipole charge, are localized at the
source. This may have important consequences for the black-hole/string correspondence principle \cite{susskind,horpol}.

\section{Isolated horizons and the correspondence principle}

The black-hole/string correspondence principle asserts that there is a smooth transition from a black hole to a string in
the limit when the string coupling is decreased \cite{susskind}.  Let us briefly discuss two subtleties which suggest that
IHs (or their nonequilibrium generalizations such as dynamical horizons \cite{ashkri2,ashkri3}) should be the most
appropriate framework for studying black hole physics in superstring theory.

For the correspondence principle to work, the entropies of the black hole and string are required to be equal for a
particular value of the string coupling constant, which ultimately means that the conserved charges of the two states
must overlap \cite{horpol}.  However, the conserved charges of the black hole (other than the dipole charge) are
typically measured at infinity (e.g. for Killing horizons), while the conserved charges of the string are localized
on the string state; to define the conserved charges of the string \emph{no reference needs to be made to infinity
at all}.  The conserved charges of the black hole should therefore \emph{not} be defined at infinity!

In addition, specification of the conserved charges of the black hole requires an \emph{a priori} knowledge of the
internal topology, e.g. typically some density is integrated over a $(D-2)$-dimensional surface with some topology
such as $S^{2}$ in four dimensions and $S^{3}$ or $S^{1}\times S^{2}$ in five dimensions.  At the transition point
when the conserved charges are equal, however, the topology of the black hole is not really important because the
spacetime loses its metric interpretation.  Therefore a framework for black holes should be employed that does not
in any way rely on the internal topology of the horizon cross sections.

As we have shown in this paper, the IH framework together with covariant phase space methods can be used to derive a
first law whereby all quantities are defined at the horizon.  In order for this derivation to work we only require that
the action be differentiable.  The conserved charges of an IH in a specific theory naturally arise after the corresponding
matter Lagrangian density is specified.  Two important properties of IHs are that the conserved charges are intrinsic to
the horizon and that there is no need to specify the topology of the horizon cross sections at any time.  IHs should
therefore be the norm rather than the exception in superstring theory.

\section*{Acknowledgements}


The author thanks Abhay Ashtekar and David Sloan for discussions, and Gary Horowitz for commenting on the manuscript.
This work was supported in part by an NSERC PDF, NSF grant PHY0854743, The George A. and Margaret M. Downsbrough Endowment
and the Eberly research funds of Penn State.



\begin{thebibliography}{99}








\bibitem{abdfklw}
A. Ashtekar, C. Beetle, O. Dreyer, S. Fairhurst, B. Krishnan, J. Lewandowski and
J. Wisniewski, \emph{Phys. Rev. Lett.} \textbf{85}, 3564 (2000)


\bibitem{hayward}
S. A. Hayward, \emph{Phys. Rev.} D \textbf{49}, 6467 (1994)

\bibitem{ashstr}
A. Ashtekar and M. Streubel, \emph{Proc. R. Soc. London} A \textbf{376}, 585 (1981)

\bibitem{leewal}
J. Lee and R. M. Wald, \emph{J. Math. Phys.} \textbf{31}, 725 (1990)

\bibitem{abr}
A. Ashtekar, L. Bombelli and O. Reula, in \emph{Analysis, Geometry and Mechanics:
200 Years After Lagrange}, edited by M. Francaviglia and D. Holm (North-Holland,
Amsterdam, 1991)

\bibitem{walzou}
R. M. Wald and A. Zoupas, \emph{Phys. Rev.} D \textbf{61}, 084027 (2000)

\bibitem{abf1}
A. Ashtekar, C. Beetle and S. Fairhurst, \emph{Class. Quantum Grav.} \textbf{16}, L1 (1999)

\bibitem{lewandowski}
J. Lewandowski \emph{Class. Quantum Grav.} \textbf{17}, L53 (2000)

\bibitem{cophor}
K. Copsey and G. T. Horowitz, \emph{Phys. Rev.} D \textbf{73}, 024015 (2006)

\bibitem{klp}
M. Korzy$\acute{\mbox{n}}$ski, J. Lewandowski and T. Pawlowski, \emph{Class. Quantum Grav.}
\textbf{22}, 2001 (2005)

\bibitem{apv}
A. Ashtekar, T. Pawlowski and C. Van Den Broeck, \emph{Class. Quantum Grav.} \textbf{24}, 625 (2007)

\bibitem{likboo}
T. Liko and I. Booth, \emph{Class. Quantum Grav.} \textbf{25}, 105020 (2008)



\bibitem{aes}
A. Ashtekar, J. Engle and D. Sloan, \emph{Class. Quantum Grav.} \textbf{25}, 095020 (2008)

\bibitem{ashslo}
A. Ashtekar and D. Sloan, \emph{Class. Quantum Grav.} \textbf{25}, 225025 (2008)

\bibitem{likslo}
T. Liko and D. Sloan, \emph{Preprint arXiv:0810.0297 [gr-qc]}

\bibitem{afk}
A. Ashtekar, S. Fairhurst and B. Krishnan, \emph{Phys. Rev.} D \textbf{62}, 104025 (2000)

\bibitem{abl}
A. Ashtekar, C. Beetle and J. Lewandowski, \emph{Phys. Rev.} D \textbf{64}, 044016 (2001)

\bibitem{caigal}
M. Cai and G. J. Galloway, \emph{Class. Quantum Grav.} \textbf{18}, 2707 (2001)

\bibitem{galloway}
G. J. Galloway, \emph{Preprint gr-qc/0608118}

\bibitem{galsch}
G. J. Galloway and R. Schoen, \emph{Commun. Math. Phys.} \textbf{266}, 571 (2006)

\bibitem{hoy}
C. Helfgott, Y. Oz and Y. Yanay, \emph{J. High Energy Phys.} \textbf{02}, 025 (2006)

\bibitem{racz}
I. R$\acute{\mbox{a}}$cz, \emph{Preprint arXiv:0806.4373 [gr-qc]}

\bibitem{abf2}
A. Ashtekar, C. Beetle and S. Fairhurst, \emph{Class. Quantum Grav.} \textbf{17}, 253 (2000)

\bibitem{gaowal}
S. Gao and R. M. Wald, \emph{Phys. Rev.} D \textbf{64}, 084020 (2001)

\bibitem{rogatko1}
M. Rogatko, \emph{Phys. Rev.} D \textbf{71}, 104004 (2005)

\bibitem{rogatko2}
M. Rogatko, \emph{Phys. Rev.} D \textbf{72}, 074008 (2005);
Erratum-Ibid. \textbf{72}, 089901 (2005)

\bibitem{ashmag}
A. Ashtekar and A. Magnon, \emph{Class. Quantum Grav.} \textbf{1}, L39 (1984)

\bibitem{ashdas}
A. Ashtekar and S. Das, \emph{Class. Quantum Grav.} \textbf{17}, L17 (2000)

\bibitem{him}
S. Hollands, A. Ishibashi and D. Marolf, \emph{Class. Quantum Grav.} \textbf{22}, 2881 (2005)

\bibitem{rogatko3}
M. Rogatko, \emph{Phys. Rev.} D \textbf{73}, 024022 (2006)

\bibitem{compere}
G. Comp$\grave{\mbox{e}}$re, \emph{Phys. Rev.} D \textbf{75}, 124020 (2007)

\bibitem{eemr1}
H. Elvang, R. Emparan, D. Mateos and H. S. Reall, \emph{Phys. Rev. Lett.} \textbf{93}, 211302 (2004)

\bibitem{eef}
H. Elvang, R. Emparan and P. Figueras, \emph{J. High Energy Phys.} \textbf{02}, 031 (2005)

\bibitem{eemr2}
H. Elvang, R. Emparan, D. Mateos and H. S. Reall, \emph{Phys. Rev.} D \textbf{71}, 024033 (2005)

\bibitem{emparan}
R. Emparan, \emph{J. High Energy Phys.} \textbf{03}, 064 (2004)

\bibitem{astrad}
D. Astefanesei and E. Radu, \emph{Phys. Rev.} D \textbf{73}, 044014 (2006)

\bibitem{susskind}
L. Susskind, \emph{Preprint hep-th/9309145}

\bibitem{horpol}
G. T. Horowitz and J. Polchinski, \emph{Phys. Rev.} D \textbf{55}, 6189 (1997)

\bibitem{ashkri2}
A. Ashtekar and B. Krishnan, \emph{Phys. Rev. Lett.} \textbf{89}, 261101 (2002)

\bibitem{ashkri3}
A. Ashtekar and B. Krishnan, \emph{Phys. Rev.} D \textbf{68}, 104030 (2003)

\end{thebibliography}
\end{document}